\documentclass[preprint,12pt,authoryear]{elsarticle}

\usepackage{amssymb}
\usepackage{amsmath}
\usepackage{multirow}
\usepackage{rotating}
\usepackage{algorithm,algorithmic}
\usepackage{xcolor}

\journal{Expert Systems With Applications}

\begin{document}

\begin{frontmatter}



\title{Consciousness-ECG Transformer for Conscious State Estimation System with Real-Time Monitoring}


\author[1]{Young-Seok Kweon}
\ead{youngseokkweon@korea.ac.kr}

\affiliation[1]{organization={Department of Brain and Cognitive Engineering, Korea University},
    city={Seoul,},
    citysep={}, 
    postcode={02841}, 
    country={Republic of Korea}}

\affiliation[2]{organization={Department of Artificial Intelligence, Korea University},
    city={Seoul,},
    citysep={}, 
    postcode={02841}, 
    country={Republic of Korea}}

\affiliation[3]{organization={Department of Biomedical Informatics Medicine, Jeju National University},
    city={Jeju,},
    citysep={}, 
    postcode={63243}, 
    country={Republic of Korea}}

    
\author[1]{Gi-Hwan Shin}
\ead{gh\_shin@korea.ac.kr}

\author[2]{Ji-Yong Kim}
\ead{jykim7@korea.ac.kr}

\author[3]{Bokyeong Ryu}
\ead{hobitmilk@jejunu.ac.kr}

\author[2]{Seong-Whan Lee\corref{cor1}}

\cortext[cor1]{Corresponding author.}
\ead{sw.lee@korea.ac.kr}

\begin{abstract}
Conscious state estimation is important in various medical settings, including sleep staging and anesthesia management, to ensure patient safety and optimize health outcomes. Traditional methods predominantly utilize electroencephalography (EEG), which faces challenges such as high sensitivity to noise and the requirement for controlled environments. In this study, we propose the consciousness-ECG transformer that leverages electrocardiography (ECG) signals for non-invasive and reliable conscious state estimation. Our approach employs a transformer with decoupled query attention to effectively capture heart rate variability features that distinguish between conscious and unconscious states. We implemented the conscious state estimation system with real-time monitoring and validated our system on datasets involving sleep staging and anesthesia level monitoring during surgeries. Experimental results demonstrate that our model outperforms baseline models, achieving accuracies of 0.877 on sleep staging and 0.880 on anesthesia level monitoring. Moreover, our model achieves the highest area under curve values of 0.786 and 0.895 on sleep staging and anesthesia level monitoring, respectively. The proposed system offers a practical and robust alternative to EEG-based methods, particularly suited for dynamic clinical environments. Our results highlight the potential of ECG-based consciousness monitoring to enhance patient safety and advance our understanding of conscious states.
\end{abstract}



\end{frontmatter}


\section{Introduction}
\label{sec:introduction}
Consciousness refers to the awareness of and response to the surrounding environment \citep{consci, lee2023lstm}. Conscious states undergo natural transitions driven by the body’s physiological requirements. For instance, to promote physical recovery, memory consolidation, and the maintenance and enhancement of cognitive function, sleep is required for part of each day \citep{sleep1, sleep2, huang2025advancing}. Conversely, conscious states can be intentionally altered for practical purposes. For example, anesthesia is administered to ensure that patients do not feel pain and remain safe during surgical procedures \citep{anesthesia1}. In both cases, individuals are in an unconscious state, which makes it crucial to distinguish between conscious and unconscious states to optimize medical interventions and advance our understanding of physiology and consciousness. However, precise and non-invasive methods to monitor conscious states remain a challenge, particularly in clinical settings where reliability and efficiency are important \citep{BIS_intra, AMS, HRSFL}. 

Monitoring conscious states is important in various medical contexts to ensure patient safety and enhance health outcomes. In surgical settings, inconsistent monitoring can lead to suboptimal anesthesia management and increase the risk of intraoperative awareness, where patients unexpectedly regain partial consciousness during surgery, causing significant distress and emotional trauma \citep{intra1, intra2, intra3}. Effective real-time monitoring allows for precise adjustments to anesthetic dosage, preventing such incidents. Beyond anesthesia, understanding conscious states during sleep holds promise for improving cognitive function and overall well-being \citep{sleep1, sleep2, AMS, HRSFL}. For instance, identifying deep sleep enables targeted memory reactivation, which enhances learning outcomes \citep{tmr1, tmr2, tmr3}. Sleep staging aids in diagnosing and treating conditions like insomnia, a disorder that can lead to serious cognitive and emotional complications \citep{insomnia1, insomnia2}. By advancing our ability to monitor and interpret conscious states, we can enhance patient safety in clinical settings and promote better health in everyday life.

Changes in the conscious state are accompanied by significant physiological and neurological transformations \citep{consci, ans1, ans2, wang2022ensemble}. During deep sleep and anesthesia, the brain and neural activity are suppressed, but during wakefulness, the brain is activated \citep{hrv_dementia, thala1, thala2}. Moreover, the balance of the autonomic nervous system (ANS) also shifts, influencing the heart rate (HR) and heart rate variability (HRV) \citep{ans1, ans2}. In wakeful states, sympathetic nervous system activity predominates, preparing the body for action by increasing HR, blood pressure, respiratory rate, and reduced HRV \citep{anesthesia_sns, sleep_sns1, sleep_sns2}. Conversely, during deep sleep or under anesthesia, parasympathetic nervous system activity takes precedence, promoting physiological restoration by slowing HR and increasing HRV \citep{sleep_hr1, sleep_hrv, anesthesia_hrv1, anesthesia_hrv2}. 

Given these physiological changes, tools that non-invasively monitor brain activity, such as electroencephalography (EEG), and those that capture ANS dynamics, like HRV derived from electrocardiography (ECG), can be employed for conscious state estimation. While EEG is widely used for assessing brain activity \cite{Kim2015Abstract, Prabhakar2020Schizo}, it faces notable challenges in clinical settings, such as high sensitivity to noise and the need for a controlled environment to obtain reliable signals \citep{eeg_noise}. Moreover, EEG is not routinely integrated into operating rooms, making its use in surgical environments potentially inconvenient, less effective, and associated with additional costs. While EEG-based methods, such as the bispectral index, are used in some operating environments, their adoption is limited by practical challenges, including cost and sensitivity to noise \citep{bis}. In contrast, ECG offers several advantages for conscious state estimation. HRV measurements from ECG reflect key changes in autonomic regulation tied to shifts in consciousness \citep{sleep_hrv, anesthesia_hrv1}. Since ECG is already an integral part of standard vital sign monitoring during surgeries, it provides greater accessibility without additional equipment. Furthermore, ECG signals are generally more robust and less susceptible to external noise than EEG, making them especially suitable for the dynamic and noise-prone environment of the operating room. Consequently, ECG stands out as a practical and effective tool for estimating conscious states, particularly in clinical scenarios where efficiency and reliability are important.

Deep learning has emerged as a powerful tool for analyzing ECG signals \citep{merdjanovska2022comprehensive}, enabling accurate and efficient classification of physiological states \citep{DKAN, REN, DRNN, DSN, SEN, zhou2021fully, liu2025net}. Recurrent neural networks (RNNs), such as long short-term memory networks (LSTMs) and gated recurrent units (GRUs), are widely used for analyzing biomedical signals \cite{Cho2021NeuroGrasp, AS, DSN}, including ECG \citep{lstm, lstm2, SEN}. Beyond RNNs, transformers have gained significant attention in recent years for their ability to capture complex temporal dependencies in sequential data \citep{transformer, ET}. Originally developed for natural language processing, transformers have been successfully adapted to biomedical signal analysis, including ECG \citep{swindae, denoiseECG, CAT, vitECG, ET}. These models have achieved remarkable performance in diverse tasks such as denoising ECG \citep{denoiseECG} and evaluation of ECG quality \citep{swindae}, demonstrating their capability to extract meaningful features from noisy and complex ECG signals. The success of transformers in these areas highlights their potential for conscious state estimation, as they can effectively model the intricate temporal patterns present in ECG data. While recurrent architectures like LSTMs and GRUs also process sequential data, they can struggle to capture long-range dependencies effectively due to issues like vanishing or exploding gradients and implicit state compression \citep{lstm1997}. In contrast, the transformer's self-attention mechanism allows it to weigh the influence of all time points within the input window \citep{transformer}. This makes them well-suited for modeling the subtle, long-range patterns in HRV for distinguishing conscious states. Building on these successes, transformers are well-positioned to tackle the challenges of modeling subtle temporal variations in ECG data that correspond to transitions in conscious states.

In this study, we proposed the consciousness-ECG transformer that consists of convolutional neural network (CNN) and transformer with decoupled query attention. This architecture enables the model to focus on the cardiac cycle, capturing subtle temporal patterns in ECG data that reflect changes in the ANS. By concentrating on these features, the model effectively estimates conscious states, such as sleep stages and anesthesia levels. Furthermore, the developed conscious state estimation system is evaluated on real-world datasets, including ECG signals collected during surgical procedures and sleep studies. These evaluations demonstrate the system's robustness and practicality, particularly in dynamic and noise-prone environments like operating rooms. This highlights its potential as a reliable tool for conscious state estimation, addressing the challenges of existing monitoring methods in both clinical and research settings.

\begin{figure*}[t!]
\centering
\includegraphics[width=\textwidth]{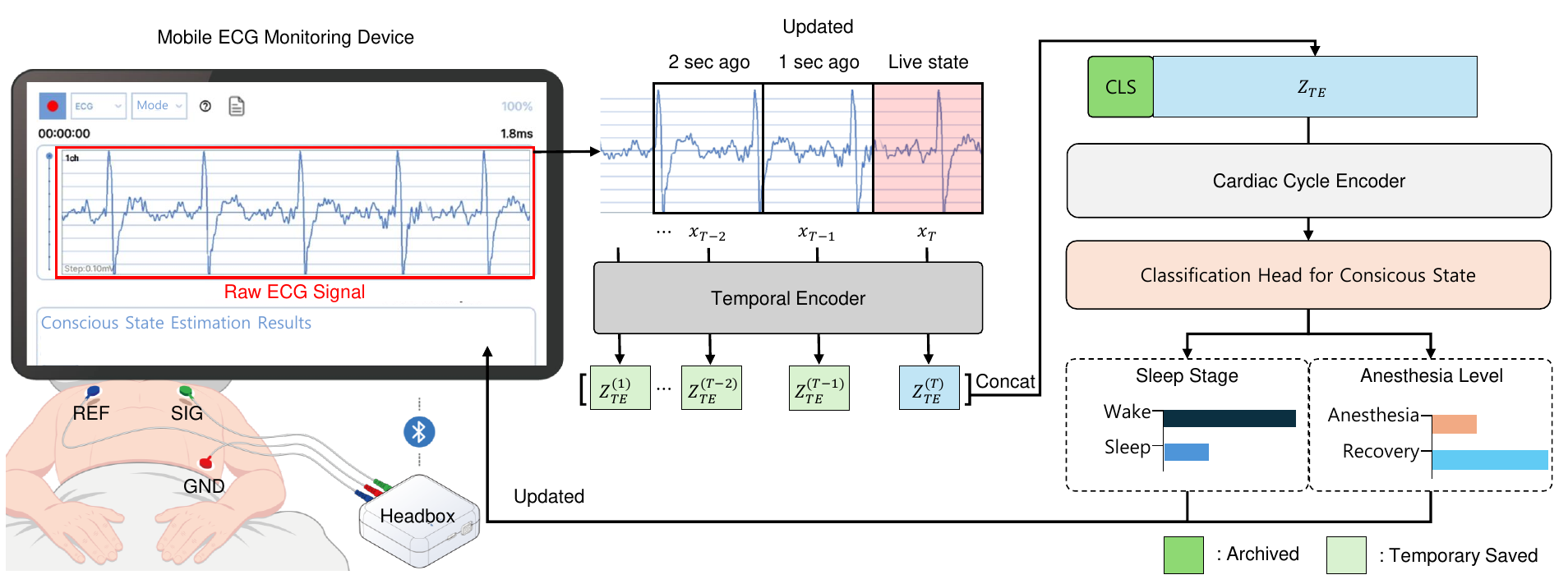}
\caption{The overall conscious state estimation system with real-time monitoring, which includes the mobile ECG monitoring device, temporal encoder, cardiac cycle encoder, and classification head. }
\label{model}
\end{figure*}

\section{Methods}

In this section, we provide an overview of the methodology for the proposed consciousness-ECG transformer and the mobile-based conscious state estimation system (Fig.~\ref{model}). First, we introduce the consciousness-ECG transformer, which consists of the temporal encoder (TE), responsible for extracting cardiac cycle features from raw ECG signals through convolutional and pooling operations, and the cardiac cycle encoder (CCE), employing a transformer architecture with decoupled query attention to capture inter-beat relationships and HRV. We briefly outline the optimization process that combines these representations for conscious state estimation. Next, we describe the mobile-based conscious state estimation system, composed of a real-time monitoring device and a mobile application designed to estimate the conscious state in real-time.

\subsection{Consciousness-ECG Transformer}

\subsubsection{Temporal Encoder}

The first stage of consciousness-ECG transformer is the TE, which transforms raw ECG time series signals into high-level feature representations. Let \(\mathrm{x} \in \mathbb{R}^{T\times F_{s}}\) be the input ECG, where \(T\) is the duration of the input ECG segment and \(F_{s}\) is the sampling rate. The goal of the TE is to produce encoded segments \(\mathbf{Z}_{TE} \in \mathbb{R}^{T \times d}\), focusing on cardiac cycles, where \(d\) is the embedding dimension. This approach allows significant memory savings, storing \(\mathbb{R}^{d}\) processed by TE instead of the raw \(\mathbb{R}^{F_{s}}\) every second, thereby reducing memory usage by a factor of \(\frac{d}{F_{s}}\).

\begin{figure*}[t!]
\centering
\includegraphics[width=0.9\textwidth]{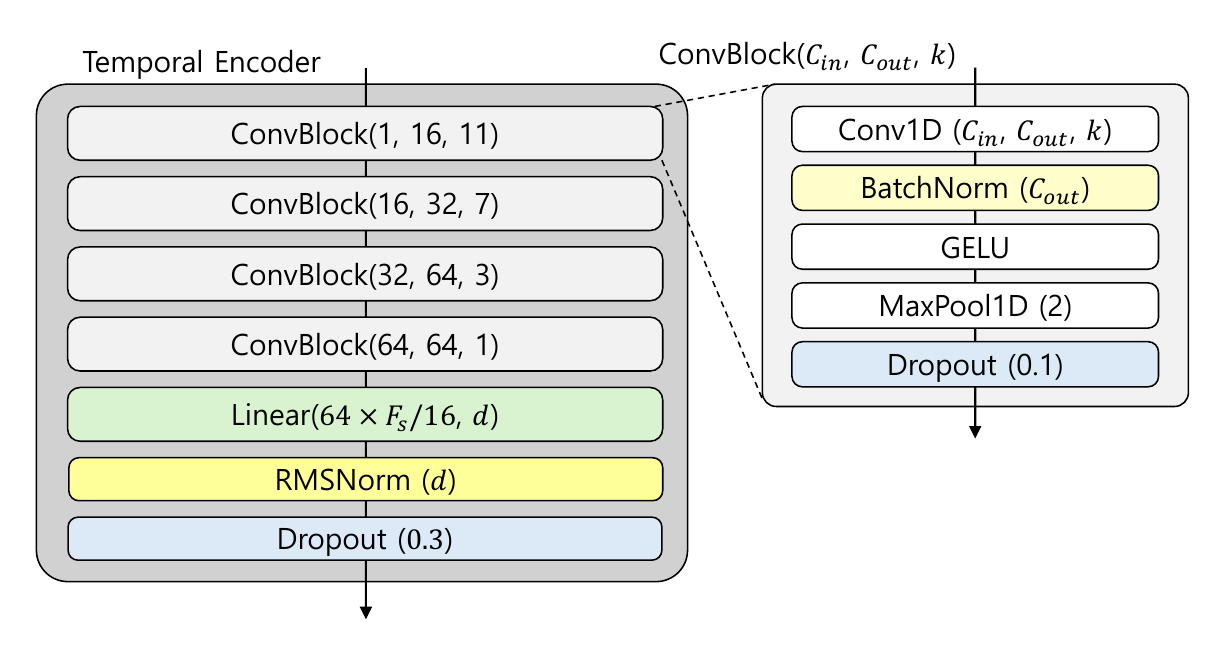}
\caption{Architecture of temporal encoder and ConvBlock.}
\label{TE}
\end{figure*}

The TE consists of multiple ConvBlock modules interleaved with max-pooling operations (Fig.~\ref{TE}). Each ConvBlock includes a one-dimensional convolution layer, batch normalization, the Gaussian error linear unit (GELU) activation, a one-dimensional max-pooling layer, and a dropout layer~\citep{gelu}. These operations reduce the temporal resolution, allowing the network to focus on essential patterns while discarding redundant information. Following these blocks, a linear layer, root mean square normalization (RMSNorm), and a dropout layer are used to further refine the features to dimension \(d\).

\subsubsection{Cardiac Cycle Encoder}
\begin{figure*}[t!]
\centerline{\includegraphics[width=0.9\columnwidth]{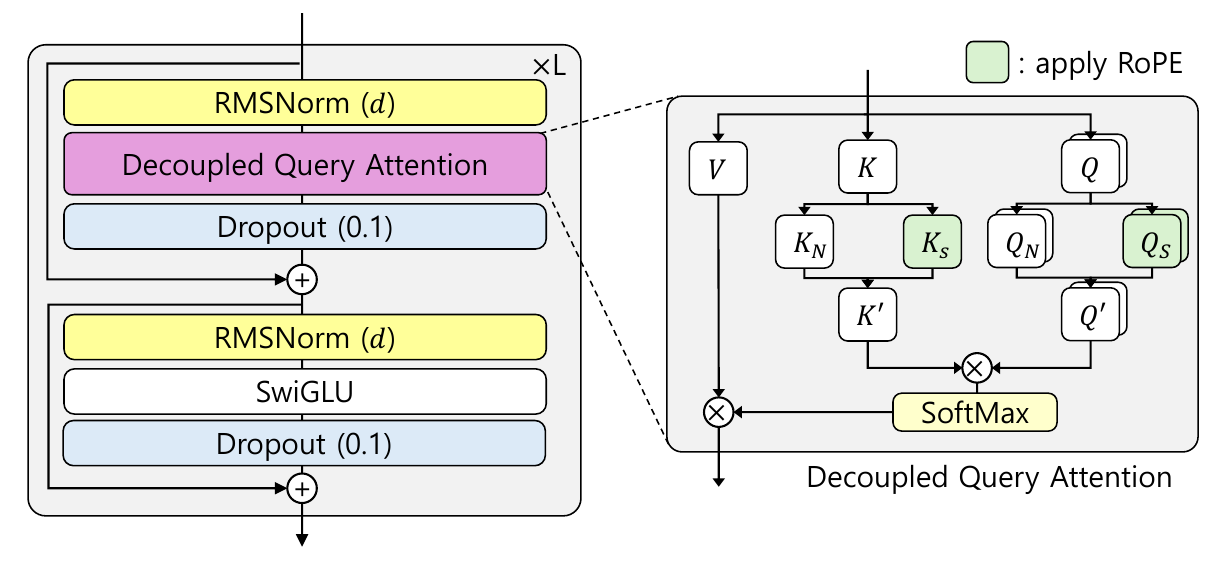}}
\caption{Architecture of cardiac cycle encoder and decoupled query attention; RoPE represents the rotary positional embedding.}
\label{CCE}
\end{figure*}

Once the TE has extracted local cardiac cycle features from the ECG, the CCE models inter-beat relationships and temporal context—essential for detecting HRV changes across conscious states. The CCE leverages a transformer-based architecture with decoupled query attention, implemented as a stack of customized transformer encoder layer units (Fig.~\ref{CCE}).

A learnable classification token, $\mathbf{z}_\mathrm{cls} \in \mathbb{R}^{1 \times d}$, is appended to the sequence prior to the transformer layers, serving as a global aggregator~\citep{vit, vitECG}:
\begin{equation}
\mathbf{Z} = \mathrm{concat}\bigl(\mathbf{z}_\mathrm{cls}, \mathbf{Z}_{\mathrm{TE}}\bigr).
\label{eq:cls}
\end{equation}

Each encoder layer processes the input $\mathbf{Z} \in \mathbb{R}^{(T + 1) \times d}$ through decoupled query attention. Decoupled query attention captures content- and position-based dependencies by grouping attention heads, so that queries in each group share the same key and value projections. Unlike conventional attention mechanisms that process features uniformly, this architecture is specifically designed to disentangle the distinct intra-beat morphological features from the inter-beat temporal dynamics inherent in ECG signals, allowing for more specialized feature extraction. In our implementation, the $H$ heads are partitioned into pairs, ensuring each key and value projection is attended by two query heads. We project $\mathbf{Z}$ into queries, keys, and values as follows:
\begin{equation}
\mathbf{Q} = \mathbf{Z}\,\mathbf{W}_\mathbf{Q},\quad
\mathbf{K} = \mathbf{Z}\,\mathbf{W}_\mathbf{K},\quad
\mathbf{V} = \mathbf{Z}\,\mathbf{W}_\mathbf{V},
\end{equation}
where $\mathbf{Q}\in\mathbb{R}^{(T+1)\times d}$ and $\mathbf{K},\mathbf{V}\in\mathbb{R}^{(T+1)\times d/2}$. To incorporate positional context efficiently, we utilized rotary positional encoding (RoPE), which applies rotations directly in the embedding space~\citep{RoPE}. We split each head’s embedding of size $d_h= d/H$ into two halves and apply RoPE only to the first half:
\begin{align}
\mathbf{Q}'^{(h)} &= \mathrm{concat}\bigl(\mathrm{RoPE}(\mathbf{Q}_S^{(h)}),\, \mathbf{Q}_N^{(h)}\bigr), \\
\mathbf{K}'^{(h)} &= \mathrm{concat}\bigl(\mathrm{RoPE}(\mathbf{K}_S^{(h)}),\, \mathbf{K}_N^{(h)}\bigr),
\end{align}
where $\mathbf{Q}_S^{(h)}, \mathbf{Q}_N^{(h)} \in\mathbb{R}^{(T+1)\times d_h/2}$ denote the split halves of $\mathbf{Q}^{(h)}$, and $\mathbf{K}_S^{(h)}$, $\mathbf{K}_N^{(h)}$ likewise for $\mathbf{K}^{(h)}$. Only the first halves ($\mathbf{Q}_S^{(h)}$, $\mathbf{K}_S^{(h)}$) are subjected to rotary transformations, embedding relative positional information into the attention mechanism without introducing additional learnable parameters. The attention score for head \(h\) is computed as follows:
\begin{equation}
A^{(h)}_{ij}
= \frac{\mathbf{Q}'^{(h)}_{i}\cdot\mathbf{K}'^{(h)}_{j}}{\sqrt{d_h}}.
\end{equation}

After computing all $\mathbf{A}_{ij}^{(h)}$, a softmax is applied to obtain attention weights, and the output context vectors are aggregated as:
\begin{equation}
\mathbf{O}_i^{(h)} = \sum_{j=1}^{T+1} 
\mathrm{softmax}\left(\mathbf{A}_{ij}^{(h)}\right) 
\mathbf{V}_j^{(h)}.
\end{equation}

The final multi-head output is formed by concatenating and linearly projecting all $\mathbf{O}_i^{(h)}$ from different heads.

A swish-gated linear unit (SwiGLU) projects the attention output to a higher dimension $\mathbb{R}^{(T+1)\times 4d}$, applies the Swish activation, and then projects it back to $\mathbb{R}^{(T+1)\times d}$~\citep{swiglu}. We employ SwiGLU as its gating mechanism provides dynamic, input-dependent control over the information flow, which enhances the model's expressive power and leads to improved performance over standard feed-forward networks. We apply RMSNorm and a residual connection before each self-attention and SwiGLU block, followed by dropout to stabilize training and mitigate vanishing gradients. We stack $L$ encoder layers in the CCE to produce the global cardiac cycle features:
\begin{equation}
  \mathbf{Z}_{\mathrm{CCE}} = \mathrm{CCE}(\mathbf{Z}).
\label{eq:cce}
\end{equation}

\subsubsection{Classification Head}
The final classification head (CH) is responsible for mapping the learned representation of the cardiac cycles to the target conscious states. First, we extract the output representation corresponding to the classification token, $\mathbf{z}_\mathrm{cls}$, from the $\mathbf{Z}_{\mathrm{CCE}}$. This token's representation serves as a global summary of the entire input sequence. Subsequently, the extracted token is passed through a single linear layer to produce the final output logits:
\begin{equation}
\mathbf{\hat{y}} = \mathbf{z}_\mathrm{cls}\mathbf{W}_{\mathrm{CH}} + \mathbf{b}_{\mathrm{CH}},
\end{equation}
where $\mathbf{W}_{\mathrm{CH}}$ and $\mathbf{b}_{\mathrm{CH}}$ are the learnable weight matrix and bias vector of the classification head, respectively.

\subsubsection{Optimization}

To mitigate label noise and address class imbalance, mini-batch samples were augmented using the mix-up method \citep{mixup}. In mix-up method, a coefficient $\lambda$ is sampled from a $\operatorname{Beta}(\alpha,\alpha)$ distribution, and each input tensor and its corresponding label are linearly combined with another randomly chosen input-label pair. This creates virtual samples that are convex combinations of original inputs and their one-hot labels, promoting linear behavior between classes and reducing model overconfidence.

In our experiments, we employed the focal loss as our objective function to further tackle potential class imbalance and enhance the learning of hard-to-classify examples \citep{focal_loss}. Given the predicted probability \( p_t \) for the true class, the focal loss is defined as:
\begin{equation}
    \text{FL}(p_t) = -w_t (1 - p_t)^\gamma \log(p_t),
    \label{eq:focal_loss}
\end{equation}
where \(w_t\) is a weighting factor to handle class imbalance, and \(\gamma\) controls the degree of down-weighting applied to easily-classified samples, thus increasing the model's focus on challenging or misclassified instances. Given the number of samples in class \(t\), \(n_t\), a weighting factor is computed based on the effective number of samples as:
\begin{align}
e_t \;=\; \frac{1-\epsilon^{\,n_t}}{1-\epsilon}, \qquad 
\tilde{w}_t \;=\; \frac{1}{e_t}, \qquad
w_t \;=\; \frac{\tilde{w}_t}{\sum_{k=1}^{K} \tilde{w}_k}\;\cdot\;\Big(\sum_{k=1}^{K} n_k\Big),
\end{align}
where \(K\) is the number of classes and $\epsilon\in[0,1)$ controls smoothing. 

\begin{algorithm}[t]
\caption{Pseudo-algorithm for training}
\label{alg:pretrain}
\begin{algorithmic}  
\STATE \textbf{Input:} Training dataset $\mathbb{D} = \{X, Y\}$; \\
\hspace*{2.8em} Network parameters $\theta = \{\theta_{TE}$, $\theta_{CCE}$, $\theta_{CH}\}$; \\
\hspace*{2.8em} Hyper-parameters $\alpha, w, \gamma$
\STATE \textbf{Output:} Optimized network parameters $\theta^{*}$
\end{algorithmic}
\begin{algorithmic}[1]  
\FOR{\(t = 0\) \textbf{to} \(\text{Maximum epoch} - 1\)}
    \FORALL{$\mathrm{x}, \mathrm{y} \sim \mathbb{D}$}
        \STATE $\lambda \sim \operatorname{Beta}(\alpha,\alpha)$
        \STATE $\mathrm{x}_{m}, \mathrm{y}_{m} = \text{Mixup}(\mathrm{x}, \mathrm{y}, \lambda) $
        \STATE $\mathrm{Z}_{\mathrm{TE}} \gets \mathrm{TE}(\mathrm{x}_{m})$
        \STATE $\mathrm{Z}_{\mathrm{CCE}} \gets \mathrm{CCE}(\text{concat}(\mathrm{z}_{cls}, \mathrm{Z}_{\mathrm{TE}}))$ \hfill \(\triangleright\) Eq.~\eqref{eq:cls}
        \STATE $\mathrm{z}_{cls} \gets \mathrm{Z}_{\mathrm{CCE}}[0]$
        \STATE $\hat{\mathrm{y}} \gets \mathrm{CH}(\mathrm{z}_{cls})$ 
        \STATE $p \gets \mathrm{softmax}(\hat{\mathrm{y}})$  
        \STATE $\mathcal{L}(\theta) \gets - \sum \mathrm{y}_{m,c} w_c \bigl(1 - p_c\bigr)^\gamma \log \bigl(p_c\bigr)$ \hfill \(\triangleright\) Eq.~\eqref{eq:focal_loss}
        \STATE $\theta \gets \text{AdamW}(\theta)$ \hfill \(\triangleright\) Eq.~(\ref{eq:adamw})
    \ENDFOR
\ENDFOR
\end{algorithmic}
\end{algorithm}

We optimized model parameters using AdamW, defined as follows:
\begin{align}
    g_t      &= \nabla_{\theta}\,\mathcal{L}(\theta), \\
    m_t      &= \frac{\beta_{1}\,m_{t-1} + (1-\beta_{1})\,g_t}{1-\beta_{1}^{\,t}}, \\
    v_t      &= \frac{\beta_{2}\,v_{t-1} + (1-\beta_{2})\,g_t^{2}}{1-\beta_{2}^{\,t}}, \\
    \theta_{t+1} &= \text{AdamW}(\theta_t) = \theta_t
                  - \eta\,\frac{m_t}{\sqrt{v_t} + \varepsilon}
                  - \eta\,\lambda_{\mathrm{w}}\,\theta_t 
    \label{eq:adamw}
\end{align}
where \(\theta\) denotes the model parameters, \(\eta\) is the learning rate, and \(\mathcal{L}\) represents the focal loss. We did not apply weight decay, as the combined use of mix-up and focal loss already provided sufficient regularization.

\begin{figure*}[!t]
\centering
\includegraphics[width=\textwidth]{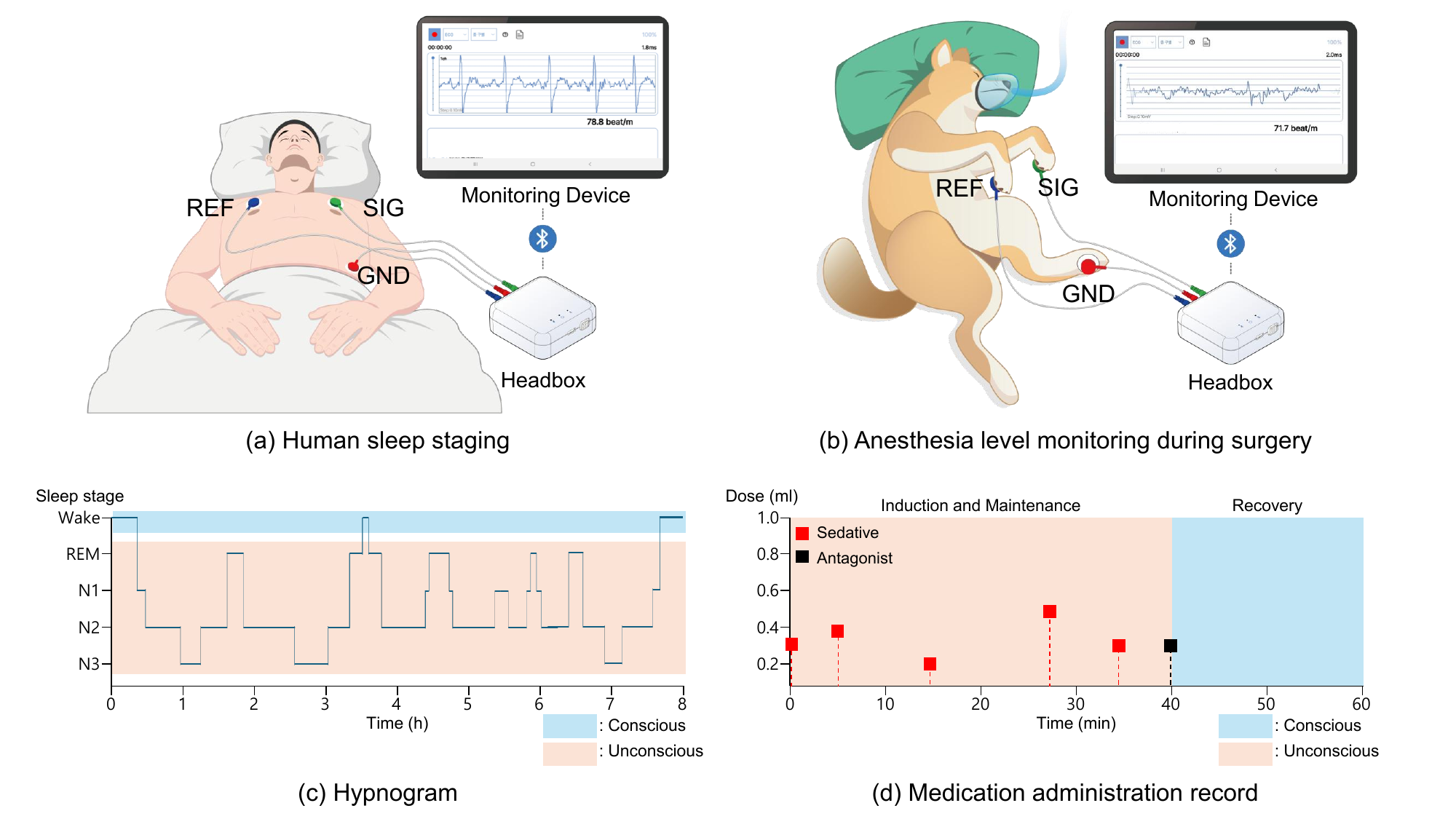}
\caption{(a) Setting of developed conscious state estimation system in overnight sleep staging. SIG, REF, and GND represent the signal, reference, and ground electrodes, respectively. (b) Setting of developed conscious state estimation system during anesthesia level monitoring. (c) Example of hypnogram from overnight sleep staging. N1, N2, and N3 represent non-rapid eye movement sleep stage 1, 2, 3, respectively and REM indicates rapid-eye movement sleep. (d) Example of medication administration record of sedative and antagonist from anesthesia level monitoring during surgery. }
\label{experiment}
\end{figure*}

\subsection{Mobile-based Conscious State Estimation System}
\subsubsection{Real-time Monitoring}

The developed real-time monitoring system consisted of a sensor, a headbox, and a monitoring device. The dimensions of the headbox are 70 mm in width, 70 mm in length, and 26 mm in height. It is certified as a medical device for animals in South Korea. The battery can last up to 13 hours on a full charge. Wireless communication is based on Bluetooth 5.0 with the SPP profile. The device operates on a single-channel system using a 3-electrode configuration (SIG, REF, and GND). As shown in Fig.~\ref{experiment}, electrodes were set to designated position for each experiment. The data sampling rate is 512 Hz. The sensor was wired to the headbox, and the measured data was transmitted to the monitoring device via Bluetooth communication in the headbox. The application installed on the monitoring device was equipped with a proposed transformer-based model, and it estimates sleep stages and anesthesia levels.

\subsubsection{Mobile-based Conscious State Estimation Application}

The mobile application was developed using Android Studio (Google LLC, Mountain View, CA, USA) and tested on a Samsung Galaxy Tab A11 (Samsung Electronics Co., Ltd., Suwon, Republic of Korea). It receives real‑time ECG data from the headbox at 512 Hz via Bluetooth 5.0, plots the live ECG signal on screen, and simultaneously stores all raw data locally. The user interface provides a mode selector to choose which conscious state to detect (anesthesia vs. recovery or wake vs. sleep), and displays the estimated state in real-time. Detected states can be shared directly from the application interface for downstream visualization.

\section{Experiments}

\subsection{Overnight Sleep Staging}

As shown in Fig.~\ref{experiment}(a), we collected ECG signals from human participants during sleep using our developed real-time monitoring system. This was measured simultaneously with the Philips Alice 6 PSG system (Philips, Amsterdam, Netherlands), and a sleep technician read the sleep stages using data from the conventional PSG system. A total of 20 participants took part in the experiment. We instructed participants to get at least 7 hours of sleep the day before the experiment, and to avoid caffeine and alcohol the day before and the day of the experiment. Participants were healthy, not suffering from any sleep-related or heart-related diseases, and not taking medications that affect sleep, heart, and blood pressure. Participants were put to bed at 10:00, the lights were turned off, and they were woken up at 6:00. The study was approved by the Korea University Institutional Review Board (KUIRB-2022-0222-03) and conducted in accordance with the Declaration of Helsinki. Informed consent was obtained from all participants.

\subsection{Anesthesia Level Monitoring during Surgery}

Our developed real-time system collected ECG signals from dogs during castration or ovariohysterectomy (OHE) with proper anesthesia (Fig.~\ref{experiment}(b)). Fifteen dogs in the shelter’s care were enrolled in the experiment. To achieve the induction of the balanced general anesthesia, we administered 5 mg/kg of ketamine, 4~mg/kg of tramadol, and 0.02~mg/kg of medetomidine intravenously. Their HR and respiratory rate during castration or OHE were manually monitored by veterinary students under the supervision of the veterinarians. To maintain the proper anesthetic state, ketamine was additionally injected intravenously with caution. Eight of the 15 dogs were male and the surgery lasted an average of 30 minutes, with an average of 7.4~mg/kg of ketamine. The female dogs were operated on for an average of 2 hours and received 75.9~mg/kg of ketamine. Four dogs were excluded from the recovery phase because excessive movement and aggression made it impossible to complete that phase. The study was approved by the Jeju National University Institutional Animal Care and Use Committee (2023-0068). 

\subsection{Preprocessing}

Table~\ref{tab:dataset} shows the details of collected dataset in sleep staging and anesthesia level monitoring. For the preprocess of collected data in sleep staging, the ECG signals were downsampled to 500 Hz and segmented into 30-second epochs. The signals were processed using a bandpass filter with a frequency range of 0.5--40~Hz to eliminate high-frequency noise and low-frequency drift. For labeling, non-rapid eye movement (NREM) sleep 1, 2 and 3 stages and rapid-eye movement (REM) were combined as the unconscious state, and wakefulness was labeled as the conscious state \citep{lee2022quantifying}.

For the anesthesia level monitoring data, similar preprocessing steps were applied. The signals were downsampled to 500 Hz and segmented into 10-second epochs. Baseline wander noise was estimated using a moving average filter and subtracted from the original signal. Subsequently, a 0.5--40~Hz bandpass filter was applied to process the signals. Additionally, signals with a signal-to-noise ratio below 10 dB were excluded from the analysis to ensure the quality of the data.

\begin{table}[t!]
\caption{Details of the collected dataset in our experiments}
\label{tab:dataset}
\begin{center}
\renewcommand{\arraystretch}{1.1}
\resizebox{0.8\textwidth}{!}{%
\begin{tabular}{c|cccc}
\hline
\textbf{Experiment}          & \textbf{\#Subjects}   & \textbf{CS}        & \textbf{US}          & \textbf{\#Total Samples}   \\ 
\hline
Sleep               & 20           & 2087      & 16366       & 18483             \\
Anesthesia          & 11           & 3001      & 7221        & 10222             \\
\hline
\multicolumn{5}{l}{CS and US represent the conscious and unconscious states, respectively.}
\end{tabular}
}
\end{center}
\end{table}

\subsection{Baselines and Experimental Settings}
In this study, we compared our model against six baselines:

\begin{itemize}
  \item DeepConvNet \citep{DCN} stacks four convolution–max-pooling blocks with batch normalization.
  \item RawECGNet \citep{REN} uses a ResNet-inspired sequence of residual CNN blocks, followed by a bidirectional gated recurrent unit over concatenated window embeddings.
  \item SleepECGNet \citep{SEN} employs a time-distributed CNN to extract features from ECG segments, which are then passed to an LSTM.
  \item DeepSleepNet \citep{DSN} combines two parallel CNN streams for time-invariant feature extraction, then models temporal dependencies with a bidirectional LSTM.  
  \item AttnSleep \citep{AS} leverages a multi-resolution CNN with adaptive feature re-calibration, followed by multi-head self-attention using causal convolutions.
  \item ECGTransform \citep{ET} integrates multi-scale convolutional layers and a channel recalibration module before modeling past and future contexts with a bidirectional transformer.
\end{itemize}

We used the published implementations for DeepConvNet \citep{DCN}, DeepSleepNet \citep{DSN}, AttnSleep \citep{AS}, and ECGTransform \citep{ET}, and re-implemented RawECGNet \citep{REN} and SleepECGNet \citep{SEN} for our evaluation. To ensure a fair comparison, all baseline models were trained and evaluated using the identical data preprocessing, augmentation, and cross-validation protocol as our proposed model.

The model's hyper-parameters and optimization settings were configured to ensure optimal performance. For both sleep staging and anesthesia level monitoring, we set the embedding dimension \(d\) to 128 and the number of stacked encoder layers \(L\) in the CCE to 2. The duration of the input ECG was set to 30 in sleep staging and 10 in anesthesia level monitoring. The learning rate was set to \(7\times10^{-5}\) in sleep staging and \(10^{-5}\) in anesthesia level monitoring. The batch size was set to 128 in sleep staging and 256 in anesthesia level monitoring to balance memory usage and training efficiency, while the focal loss parameters \(\gamma\) were initialized to \(2.25\) in sleep staging and \(2.0\) in anesthesia level monitoring, to handle potential class imbalances and emphasize hard-to-classify examples. Also the \(\epsilon\) was set to \(1-10^{-2}\) in sleep staging and \(1-5\times10^{-4}\) in anesthesia level monitoring. Training was conducted for \(100\) epochs and a coefficient \(\alpha\) for mix-up is 0.3 in sleep staging and 0.5 in anesthesia. The optimization process was completed by running the model for the maximum number of epochs. The weight decay in both sleep staging and anesthesia level monitoring was set to 0 ($\lambda_{\mathrm{w}}=0$). Hyper-parameters were selected via grid search on the first cross-validation fold for each task and model, then fixed for all remaining folds.

\subsection{Evaluation}
The models for sleep staging and anesthesia level monitoring were trained and evaluated independently on their respective datasets. To evaluate our methods in sleep staging, we applied a leave‑one‑subject‑out (LOSO) strategy: in each fold, every epoch from one entire subject was held out for testing, while the remaining subjects were split epoch‑wise into training subsets. For anesthesia‑level monitoring, the 11 subjects were first partitioned into five non‑overlapping subgroups; four groups contained two subjects and one contained three. Evaluation proceeded using a LOSO strategy: in fold \(k\) the full recordings of the \(k\)‑th subgroup made up the test set, and the union of the remaining subgroups was used as the training set.

We exploited two complementary families of metrics: overall and diagnostic. Overall metrics summarize the global agreement between predictions and ground truth across two classes: accuracy (ACC), macro‑averaged F1 (MF1), and Cohen’s kappa (\(\kappa\)). These overall metrics are computed as follows:
 \begin{align}
\text{MF1} &=\frac{\sum_{i=1}^{2}\text{F1}_{i}}{2},\\
\text{ACC} &=\frac{\text{TP}+\text{TN}}{M},\\
\kappa &=\frac{\text{ACC}-P_{e}}{1-P_{e}},
\end{align}
where \(P_{e}\) is the chance agreement that is computed based on marginal totals in the confusion matrix; \(M\) is the total number of evaluated epochs; and TP and TN are the true positives and true negatives, respectively.

Diagnostic metrics—specificity (Sp), sensitivity (Se), positive predictive value (PPV), and negative predictive value (NPV)—characterize the model’s performance when distinguishing epochs belonging to the conscious state (wake in sleep staging and recovery in anesthesia level monitoring) from unconscious state (NREM and REM in sleep staging and anesthesia in anesthesia level monitoring). All these indicators are computed once from the single confusion matrix obtained by pooling the predictions of every fold \citep{AS, DSN}. Pooling predictions from all folds enables more robust estimation of metrics by aggregating results across independent test subsets. As our method yields a single point estimate for each metric, a statistical test for comparison is not directly applicable.

To assess discrimination at every decision threshold, we drew the receiver operating characteristic (ROC) curves based on the posterior probabilities generated by the model. For each binary task the predictions of all folds were pooled into a single vector of scores and labels; the Se was then plotted against the 1-Sp as the threshold varied from~$1$ to~$0$. Performance was summarized by the area under the ROC curve (AUC), interpreted as the probability that a randomly chosen positive epoch will receive a higher score than a randomly chosen negative one. To quantify the system's real-time performance, latency was measured as the time interval from the arrival of a complete ECG epoch via Bluetooth to the display of the corresponding classification result on the screen. 

\begin{table*}[!t]
\caption{Comparison of Conscious State Estimation Performance during Sleep and Anesthesia}
\label{tab:model_performance_baseline}
\begin{center}
\renewcommand{\arraystretch}{1.2}
\resizebox{\columnwidth}{!}{%
\begin{tabular}{c|c|ccc|cccc}
\hline
                &        & \multicolumn{3}{c|}{\textbf{Overall Metrics}} & \multicolumn{4}{c}{\textbf{Diagnostic Metrics}}\\
\hline
Classification  & Model & ACC   & MF1   & $\kappa$ & Sp    & Se    & PPV    & NPV    \\
\hline
                & DeepConvNet   & 0.759         & 0.529         & 0.069         & 0.268         & 0.821         & 0.898         & 0.160 \\
                & RawECGNet     & 0.556         & 0.452         & 0.040         & 0.534         & 0.559         & 0.904         & 0.134 \\
                & SleepECGNet   & 0.671         & 0.546         & 0.163         & \textbf{0.644}& 0.675         & \textbf{0.937}& 0.202 \\
Sleep           & DeepSleepNet  & 0.666         & 0.498         & 0.051         & 0.388         & 0.702         & 0.900         & 0.142 \\
                & AttnSleep     & 0.536         & 0.443         & 0.041         & 0.568         & 0.531         & 0.906         & 0.134 \\
                & ECGTransform  & 0.805         & 0.505         & 0.010         & 0.118         & 0.892         & 0.888         & 0.123 \\
                & Ours          & \textbf{0.877}& \textbf{0.691}& \textbf{0.381}& 0.447         & \textbf{0.931}& 0.930         & \textbf{0.454} \\
\hline
                & DeepConvNet  & 0.574         & 0.550         & 0.130         & 0.584         & 0.570         & 0.767         & 0.361  \\
                & RawECGNet    & 0.634         & 0.498         & 0.011         & 0.193         & 0.817         & 0.709         & 0.305  \\
                & SleepECGNet  & 0.773         & 0.729         & 0.458         & 0.630         & 0.832         & 0.844         & 0.609  \\
Anesthesia      & DeepSleepNet & 0.666         & 0.650         & 0.329         & \textbf{0.773}& 0.622         & 0.868         & 0.459  \\
                & AttnSleep    & 0.786         & 0.748         & 0.496         & 0.674         & 0.833         & 0.860         & 0.626  \\
                & ECGTransform & 0.704         & 0.579         & 0.178         & 0.269         & 0.885         & 0.745         & 0.494  \\
                & Ours         & \textbf{0.880}& \textbf{0.845}& \textbf{0.691}& 0.691         & \textbf{0.958}& \textbf{0.882}& \textbf{0.873}\\
\hline
\multicolumn{9}{l}{\textbf{Bold} indicates best performance among models}
\end{tabular}
}
\end{center}
\end{table*}

\section{Results}

\begin{figure*}[!t]
\centering
\includegraphics[width=0.7\textwidth]{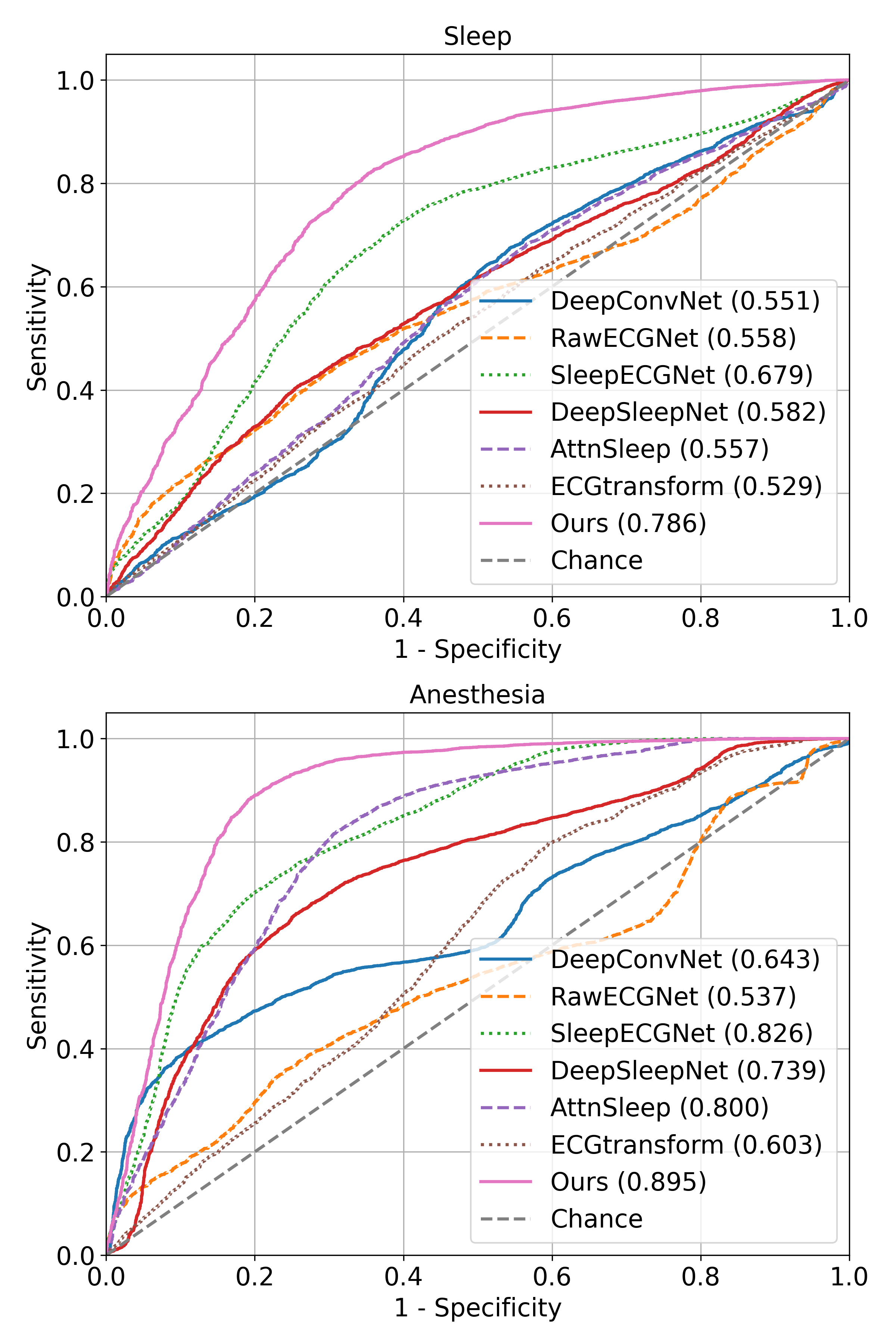}
\caption{Comparison of ROC curves and AUC values across various methods.}
\label{anesthesia_roc}
\end{figure*}

\subsection{Performance of Conscious State Estimation}
Table~\ref{tab:model_performance_baseline} reports a comprehensive comparison of our consciousness-ECG transformer against six baseline architectures on both sleep staging and anesthesia monitoring. In sleep staging, consciousness-ECG transformer achieves an accuracy of 0.877, MF1 of 0.691 and Cohen’s \(\kappa\) of 0.381, exceeding the next best model (ECGTransform) by 0.072 in accuracy and by 0.186 in \(\kappa\). For anesthesia level monitoring, consciousness-ECG transformer attains an accuracy of 0.880, MF1 of 0.845 and \(\kappa\) of 0.691—outperforming the strongest baseline (AttnSleep) by 0.094 in accuracy and by 0.195 in \(\kappa\). Notably, our method also delivers a balanced diagnostic profile, with Se and PPV both above 0.93 under anesthesia, while improving Sp and NPV by over 0.020 compared to the best baseline. Our consciousness-ECG transformer is the only model that sustains \(\kappa\) over 0.38 in sleep and \(\kappa\) over 0.69 in anesthesia using an identical method, underscoring its capacity to generalize across disparate autonomic regimes. These results demonstrate that the proposed decoupled‑query attention architecture not only yields higher aggregate accuracy but also achieves a more clinically reliable trade‑off between false positives and false negatives. Such robustness across both sleep and anesthesia regimes establishes consciousness-ECG transformer as a strong candidate for real‑time conscious‑state monitoring.

\begin{figure*}[!t]
\centering
\includegraphics[width=0.8\textwidth]{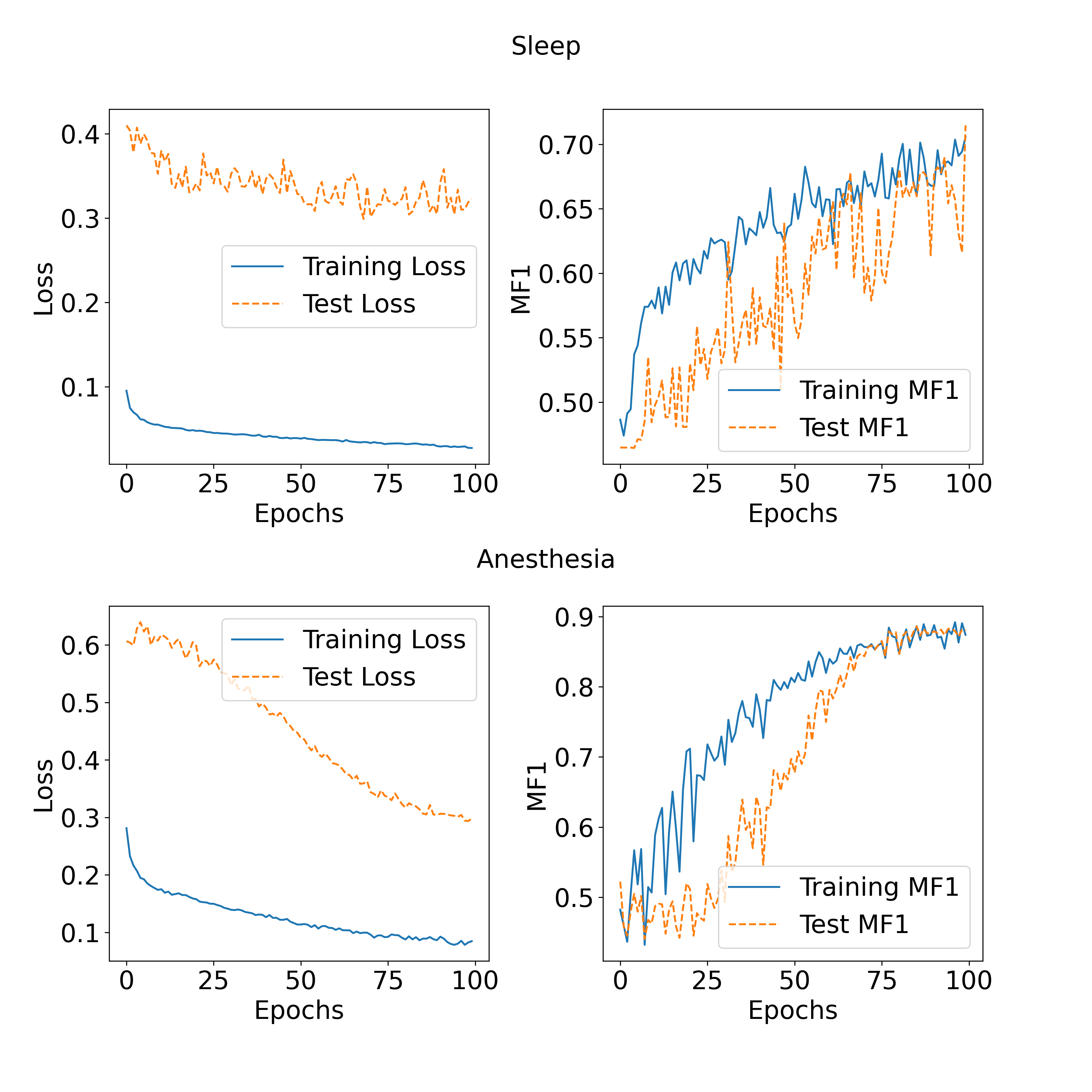} 
\caption{Training and testing loss and MF1 comparison for a random fold (i.e. fold 4 in sleep and fold 9 in anesthesia).}
\label{loss}
\end{figure*}

Figure~\ref{anesthesia_roc} presents the ROC curves and corresponding AUC values for all compared methods. In sleep staging, AUC spans from 0.529 (ECGTransform) to 0.786 (Ours), with our model surpassing the next-best baseline (SleepECGNet, 0.679) by 0.107. Under anesthesia, AUC ranges from 0.537 (RawECGNet) to 0.895 (Ours), marking an improvement of 0.069 over SleepECGNet (0.826). These results demonstrate that conscious-ECG transformer consistently achieves the highest discriminative performance across operating thresholds. 

Figure~\ref{loss} presents the training and validation loss curves for the conscious-ECG transformer over 100 epochs. The smooth decline of both curves and the narrow gap between training and validation losses confirm stable convergence without overfitting. Notably, loss plateaus around epoch 60, indicating the model has effectively learned salient ECG features by this point.

\begin{figure*}[!t]
\centering
\includegraphics[width=0.8\textwidth]{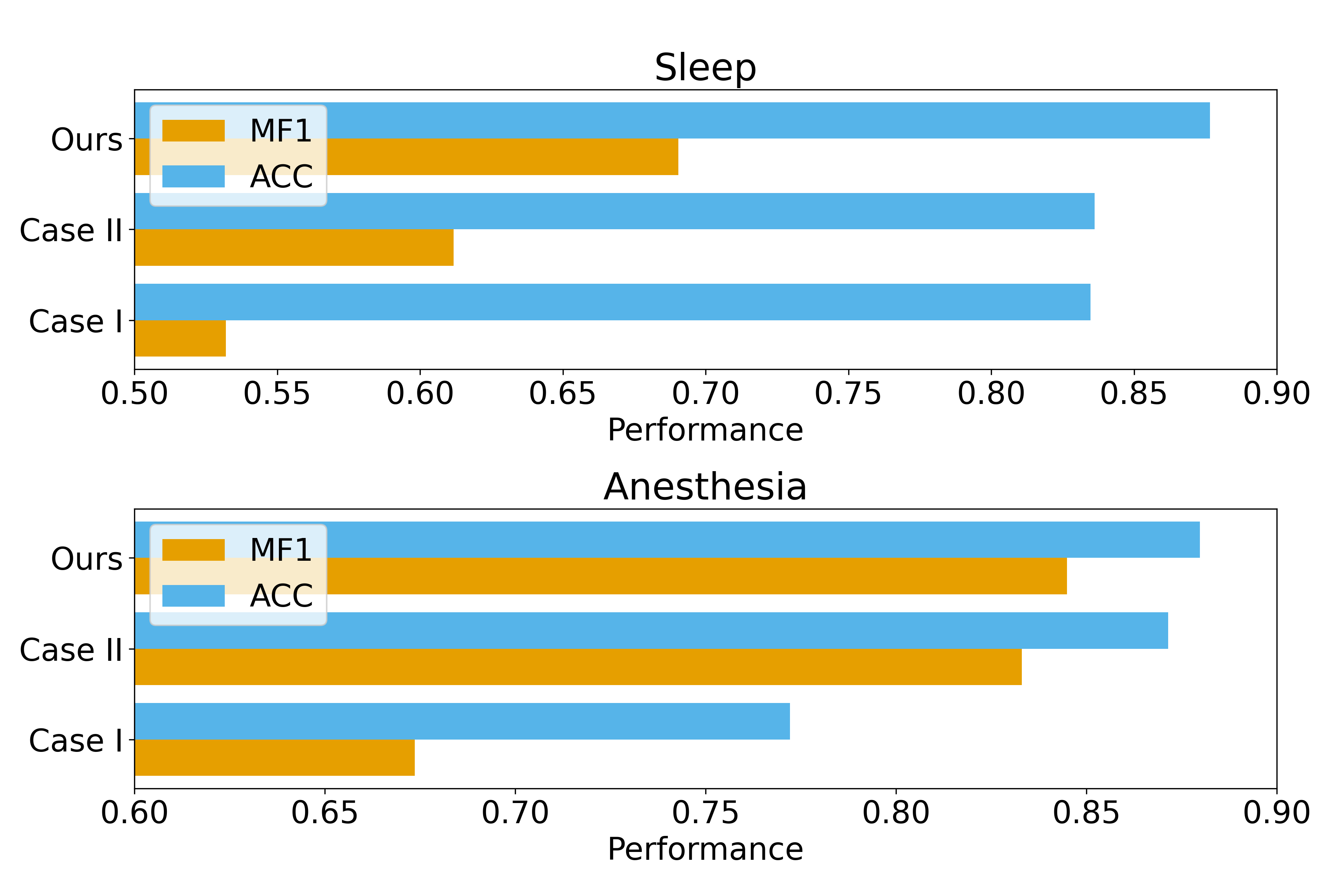}
\caption{Comparison of overall ACC and MF1 for ablation study.}
\label{bar}
\end{figure*}

\subsection{Ablation Study}

In this study, we proposed a novel feature extractor, which consists of transformer with decoupled query attention, for cardiac cycle based salient feature representation. To verify the effects of these components, we conducted ablation experiments with two variants. 
\begin{itemize}
\item Case I: We constructed a model without the CCE and, hence, the model is composed of only TE. 
\item Case II: For this case, we exploited the group query attention, omitting the decoupled attention mechanism. 
\item Consciousness-ECG Transformer: We used our proposed method that is composed of TE and CCE with decoupled attention mechanism.
\end{itemize}

As summarized in Figure~\ref{bar}, Case I exhibits the lowest accuracy (Sleep: 0.835, Anesthesia: 0.772) and MF1 (Sleep: 0.532, Anesthesia: 0.674), confirming that temporal features alone are insufficient. Introducing group query attention (Case II) yields a marked improvement in both accuracy (Sleep: 0.836, Anesthesia: 0.871) and MF1 (Sleep: 0.612, Anesthesia: 0.833), demonstrating the value of modeling inter-beat relationships. Finally, our consciousness-ECG transformer achieves the best accuracy (Sleep: 0.877, Anesthesia: 0.880) and MF1 (Sleep: 0.691, Anesthesia: 0.845)—a 0.079 and 0.012 MF1 improvement over Case II, respectively—validating the effectiveness of decoupled query attention.

\begin{figure*}[!t]
\centering
\includegraphics[width=0.8\textwidth]{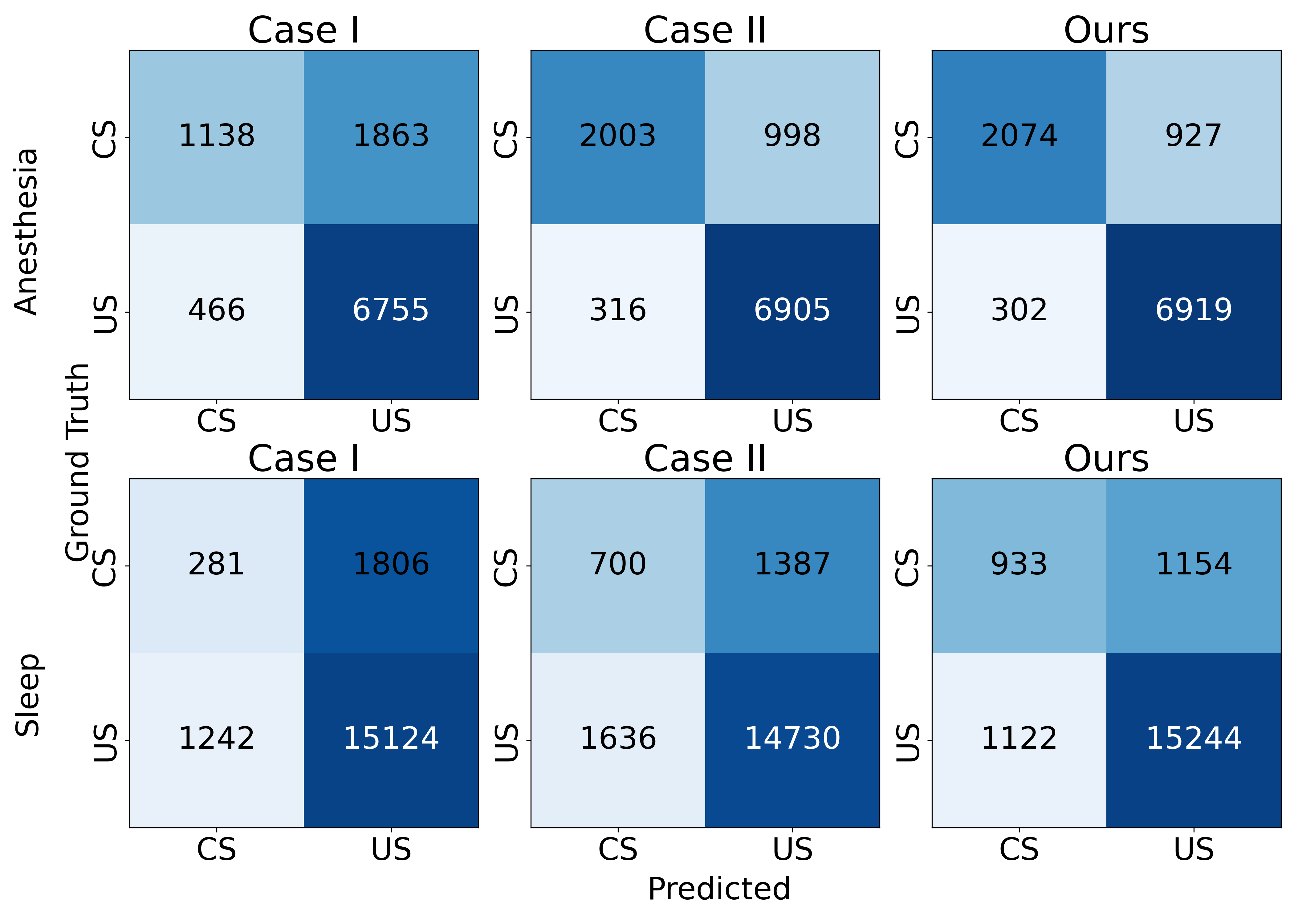}
\caption{Comparison of confusion matrices based on different cases.}
\label{cm}
\end{figure*}

Figure~\ref{cm} shows the corresponding normalized confusion matrices. In Case I most off‐diagonal entries are large, indicating frequent misclassifications. Case II reduces these errors but still confuses unconscious states as conscious in about 15\% of samples. In contrast, our decoupled-attention model yields the strongest diagonal dominance (over 90\% true‐positive rates) with minimal false positives and negatives, demonstrating robust separation of conscious versus unconscious epochs.

\section{Discussion}
Our consciousness-ECG transformer consistently outperformed six baselines on both sleep staging and anesthesia level monitoring tasks (Table~\ref{tab:model_performance_baseline}). In sleep staging, our model achieved an accuracy of 0.877, an MF1 of 0.691, and Cohen’s \(\kappa\) of 0.381. In anesthesia level monitoring, it reached an accuracy of 0.880, an MF1 of 0.845, and \(\kappa\) of 0.691. ROC analysis showed superior discrimination, with AUC values of 0.786 in sleep and 0.895 in anesthesia (Fig.~\ref{anesthesia_roc}). Ablation studies confirmed that both the TE and the decoupled‑query attention mechanism are necessary: Case I (no CCE) reduced MF1 from 0.691 to 0.532, while replacing decoupled attention with grouped attention reduced MF1 from 0.691 to 0.612 in sleep staging (Fig.~\ref{bar}). Latency of system was average 317.22 ms. These results underscore the robustness across autonomic regimes.

\subsection{Impact of Decoupled Query Attention}
The decoupled‑query attention mechanism enables the model to disentangle intra‑beat morphology (P, QRS, T waves) from inter‑beat temporal dynamics, by pairing query heads with shared key and value projections. Without RoPE, attention heads can focus on absolute signal shape—capturing subtle PQRST characteristics—while the RoPE‑augmented heads emphasize relative R–R interval variations linked to HRV, which are essential for estimating conscious states \citep{HRV_sedation,HRV_sedationmonitor,HRV_sedationmonitor2, HRV_sleep, Bae}. This selective application of RoPE stands in contrast to architectures like RoFormer, where its universal application may not optimally separate timing from morphology \citep{RoPE}. Similarly, our approach differs from the more monolithic attention in models like ECGTransform or the efficiency-focused windowing strategies of Swin-based models, as our novelty lies in the internal, physiology-aware design of the attention head itself \citep{swindae, ET}. Our ablation (Case II vs.\ Ours) shows a 0.079 increase in MF1 when using decoupled attention, demonstrating that preserving both spatial and sequential contexts is critical for conscious state estimation. 

\subsection{Clinical Applicability}

The proposed conscious state estimation system has significant potential for clinical applications. No additional EEG hardware is required, reducing setup complexity and cost while improving patient comfort. Continuous real‑time monitoring on a tablet demonstrated up to 13 hours of battery life, low-latency state updates, and immediate alerting of consciousness transitions. Our approach has the potential to decrease the incidence of intraoperative awareness, optimize anesthetic dosing, and streamline sleep disorder diagnosis. The user interface’s mode selector and one‑tap sharing further enable downstream visualization and integration with electronic health records, supporting evidence‑based decision making in dynamic clinical environments.

In the context of both sleep staging and anesthesia level monitoring, the system offers a robust and practical solution for operating room environments and sleep environments, which could aid in diagnosing and managing sleep disorders such as insomnia and sleep apnea. By providing real-time feedback on conscious states during sleep, the system could also support interventions such as targeted memory reactivation, enhancing cognitive function and learning outcomes \citep{tmr1, tmr2, tmr3}. The reliance on ECG, which is already a standard component of vital sign monitoring during surgeries, eliminates the need for additional equipment, reducing setup complexity and associated costs. Real-time estimation of anesthesia levels could enhance patient safety by minimizing risks such as intraoperative awareness and over-sedation, thereby improving overall surgical outcomes \citep{intra1, intra2, intra3}. The system's resistance to noise makes it suited to dynamic and challenging surgical settings or unpredicted sleep environments.

\subsection{Limitations and Future Work}
Despite these promising results, our study has limitations. First, the TE’s fixed receptive field assumes a normal heart rate range; extreme bradycardia or tachycardia may misalign cycle representations. Second, real-time inference on resource-constrained devices requires further optimization of model size and computational load. Third, our validation was only conducted on custom-collected datasets. While they reflect real-world clinical scenarios, evaluation on broader public datasets and human anesthesia datasets is necessary to fully assess the model's generalization capabilities.

Future work will (\textit{i}) explore dynamic receptive-field adaptation based on instantaneous HR, (\textit{ii}) conduct large-scale clinical trials and further validation across diverse public datasets and human anesthesia experiments to establish the model's generalizability, and (\textit{iii}) develop integrated explainability tools that visualize attention maps in real-time to aid clinician interpretation. 

\section{Conclusion}
In this study, we presented consciousness-ECG transformer for conscious state estimation system with real-time monitoring. Our model demonstrated superior performance over traditional baselines in both sleep staging and anesthesia level monitoring, highlighting its robustness and clinical applicability. This ECG-based system offers a practical, non-invasive alternative to EEG-based methods, enhancing patient monitoring and safety in dynamic medical environments.




\bibliographystyle{elsarticle-harv} 





\section*{Acknowledgements}
This work was partly supported by Institute of Information \& Communications Technology Planning \& Evaluation (IITP) grant funded by the Korea government (MSIT) (IITP-2025-RS-2024-00436857, ITRC (Information Technology Research Center) and No. RS-2019-II190079, Artificial Intelligence  Graduate  School  Program  (Korea University)) and the National Research Foundation of Korea (NRF) grant funded by the MSIT (No. 2022-2-00975, MetaSkin: Developing Next-generation Neurohaptic Interface Technology that enables Communication and Control in Metaverse by Skin Touch).

\bibliography{references}

\end{document}